\pdfoutput=1

\documentclass{raa}

\usepackage{mathptmx}
\usepackage{ae,aecompl}

\usepackage{graphicx}	
\usepackage{amsmath}	
\usepackage{amssymb}	
\usepackage{pifont}
\usepackage{txfonts}
\usepackage{url}
\usepackage{subfigure}
\usepackage{longtable}
\usepackage{lscape}
\usepackage{multirow}
\usepackage{color}
\usepackage{textcomp}
\usepackage{natbib}
\usepackage[colorlinks,linkcolor=red,anchorcolor=green,citecolor=blue]{hyperref}
\usepackage{threeparttable}

\begin{document}

\title{Statistical Properties of X-ray Bursts from SGR J1935+2154 detected by Insight-HXMT}


   \volnopage{Vol.0 (2021) No.0, 000--000}      
   \setcounter{page}{1}          
   \author{Wen-Long Zhang\inst{1},
      Xiu-Juan Li\inst{2},
      Yu-Peng Yang\inst{1},
      Shuang-Xi Yi\inst{1},
      Cheng-Kui Li\inst{3},
      Qing-Wen Tang\inst{4},
      Ying Qin\inst{5}
      and Fa-Yin~Wang\inst{6}
   }

   \institute{\inst{1}School of Physics and Physical Engineering, Qufu Normal University, Qufu 273165, China; {\it yisx2015@qfnu.edu.cn} \\
    \inst{2}School of Cyber Science and Engineering, Qufu Normal University, Qufu 273165, China;\\
    \inst{3}Key Laboratory of Particle Astrophysics, Institute of High Energy Physics, Chinese Academy of Sciences, Beijing 100049, China; {\it lick@ihep.ac.cn}\\
    \inst{4}Department of Physics, School of Physics and Materials Science, Nanchang University, Nanchang 330031, China; {\it qwtang@ncu.edu.cn}\\
    \inst{5}Department of Physics, Anhui Normal University, Wuhu 241000, China;\\
    \inst{6}School of Astronomy and Space Science, Nanjing University, Nanjing 210023, China}


\abstract{As one class of the most important objects in the universe, magnetars can produce a lot of different frequency bursts  including X-ray bursts. In \cite{2022ApJS..260...24C}, 75 X-ray bursts produced by magnetar SGR J1935+2154 during an active period in 2020 are published, including the duration and net photon counts of each burst, and waiting time based on the trigger time difference. In this paper, we utilize the power-law model, $dN(x)/dx\propto (x+x_0)^{-\alpha_x}$, to fit the cumulative distributions of these parameters. It can be found that all the cumulative distributions can be well fitted, which can be interpreted by a self-organizing criticality theory. Furthermore, we check whether this phenomenon still exist in different energy bands and find that there is no obvious evolution. These findings further confirm that the X-ray bursts from magnetars are likely to be generated by some self-organizing critical process, which can be explained by a possible magnetic reconnection scenario in magnetars.
\keywords{Magnetar; X-Ray bursts; SOC-system} }

   \authorrunning{Zhang et al.}                     
   \titlerunning{Statistical Properties of X-ray Bursts from SGR J1935+2154 detected by Insight-HXMT}       

   \maketitle


\section{Introduction}    
\label{sect:intro}

Magnetars are a group of isolated neutron stars, which have extremely powerful magnetic fields \citep{1995MNRAS.275..255T}. The dissipation of the magnetic field provides energy for magnetar. They are characterized by strong variability on several time scales and exhibit large variations across the electromagnetic spectrum, especially in X-ray and soft gamma ray energy bands, ranging from a few milliseconds to a month (\citealp{2017ARA&A..55..261K}). SGR J1935+2154 was discovered in 2014 when Swift-BAT (Burst Alert Telescope) was triggered by a short burst from Galactic plane \citep{2014GCN.16520....1S}. Based on a full multi-wavelength radio study of continuous and persistent emission of SGR J1935+2154 reported by \cite{2018ApJ...852...54K}, \citet{2014GCN.16533....1G} found that it was associated with supernova remnant SNR G57.2+0.8.
\cite{2020ApJ...905...99Z} corrected the distance from SGR J1935+2154 to SNR G57.2+0.8 and found that the distance between them may only be 6.6 $\pm$ 0.7 kpc, which is closer than the previous hypothesis of 10 kpc.

Since 27 April 2020, SGR J1935+2154 entered a new period of activity, in which the magnetar produced several outbursts, including fast radio burst (FRB) 200428 \citep{2020Natur.587...54C,2020Natur.587...59B}. The observation of FRB 200428 firstly confirmed the prediction that magnetar may be one of the candidates of the origin of FRBs \citep{2021MNRAS.503.5367B,2022arXiv221006547C}. In the days and months following FRB 200428, several telescopes made sustained observations. However, there is no any significant single radio pulse \citep{2021MNRAS.503.5367B}. Notably, the Insight-HXMT satellite detected a non-thermal X-ray burst associated with FRB 200428 and then it was identified as the emission from SGR J1935+2154 \citep{2021NatAs...5..378L}. Moreover, Insight-HXMT has detected 75 X-ray bursts (1- 250 keV) from SGR J1935+2154 after FRB 200428 and found that one of them shows a similar peak spectral energy from the X-ray burst associated with FRB 200428 \citep{2022ApJS..260...24C,2022ApJS..260...25C}. They found that the cumulative distribution of the fluence of all 75 bursts can be well fitted by a power-law with a index of $0.764 \pm 0.004$, which is consistent with the result of SGR J1935+2154 bursts by Fermi/GBM reported by \cite{2020ApJ...893..156L}.

Studies show that X-ray events often occur in some high-energy celestial burst events, e.g. X-ray flares from X-ray binaries or other systems, especially in the nearest star, the Sun (\citealp{2011soca.book.....A,2011LRSP....8....6S}), the most violent celestial explosions known in the universe: gamma-ray bursts (GRBs; \citealp{2005Sci...309.1833B,2006ApJ...641.1010F,2006ApJ...642..354Z,2016ApJS..224...20Y,2017ApJ...844...79Y,2016ApJ...832..161M,2021ApJ...922..255T,2023ApJS..265...56L}), some active galactic nuclei events (AGNs; \citealp{1984ARA&A..22..471R,2018ApJ...864..164Y}), and the tidal disruption events (TDE), such as Swift J1644+57 (\citealp{2011Natur.476..421B,2011Sci...333..203B,2020RAA....20...17Z}).

As is well known that the solar X-ray flares could be originated from the magnetic reconnection process, \cite{2013NatPh...9..465W} found that GRB X-ray flares show the similar power-law distributions like solar flares in terms of the waiting time, energy and duration, respectively. It is believed that both of X-ray flares from GRBs and solar flares may also have the same physical mechanism. Because they can also be estimated in a self-organized criticality (SOC) system (\citealp{1987PhRvL..59..381B,1989JGR....9415635B,1991ApJ...380L..89L,2011LRSP....8....6S,2011soca.book.....A}).

In addition, the same power-law distributions of X-ray flares or bursts have been found in other systems, for example, some black hole binary systems (\citealp{2015ApJS..216....8W,2018ApJ...864..164Y}), type I X-ray bursts originated from X-ray binary systems with low-mass (\citealp{2017MNRAS.471.2517W}), some repeating FRBs (\citealp{2017MNRAS.471.2517W,2020ApJ...893...44Z,2023arXiv230206802W}), and soft gamma repeaters \citep[e.g.][]{1996Natur.382..518C,1999ApJ...526L..93G,2000ApJ...532L.121G,2020MNRAS.491.1498C}.
The similar distributions for parameters of X-ray flares from supergiant fast X-ray transients are also reported by \cite{2016MNRAS.457.3693S} and \cite{2022RAA....22f5012Z}. Their studies indicated that the power-law characters for the supergiant fast X-ray transients are an important evidence of SOC.

The SOC behaviors of high energy celestial bodies indicate that it is necessary to systematically analyze the properties to further study their physical origin. In this work, we study the statistical properties of 75 bursts generated from SGR J1935+2154 reported by \citet{2022ApJS..260...24C}, including the durations, waiting times, and net photon counts. In Section 2, we present the sample selection and data analysis methods. Our main results and discussion are shown in Section 3. Finally, our conclusions are given in Section 4.

\section{Data and Statistical analyses}
\label{sect:data}

From 2020-04-28T07:14:51 UTC to 2020-06-01T00:00:01 UTC, Insight-HXMT conducted a 33-day ToO observation of SGR J1935+2154 (total span 2851.2 ks), during which several hundred short X-ray bursts were emitted from SGR J1935+2154. The detailed observation time is listed in Table 4 reported by \citet{2022ApJS..260...24C}. They provide a comprehensive monitoring of the evolution of burst activity from SGR J1935+2154 with high temporal resolution and sensitivity over a very broad energy range (1-250 keV). In this work, we study three parameters, including duration, waiting time and net photon counts. The waiting time is defined as the difference between the beginning time of the $i + 1th$ and $ith$ burst, that is, $T_{waiting}=T_{start,i+1}-T_{start,i}$ \citep{2016ApJS..224...20Y}.

In general, the differential distribution could be described by a threshold power-law distribution with the following Equation
\begin{equation}
\frac{dN}{dx}\propto (x+x_{0})^{-\alpha_{x}}, x_{1}\le x\le x_{2}.
\label{diff_cum}
\end{equation}

The cumulative distribution of such explosive events can be written as the integral of the whole number of events exceed a given value x, so the cumulative distribution function corresponding to Eq.~(\ref{diff_cum}) can be expressed as ($\alpha_x\neq1$) \citep{2015ApJ...814...19A}
\begin{equation}
N_{\mathrm{cum}} (>x) = 1+(N_{\rm env}-1)\times \left(\frac{(x_{2}+x_{0})^{1-\alpha_x}-(x+x_{0})^{1-\alpha_x}}{(x_{2}+x_{0})^{1-\alpha_x}-(x_{1}+x_{0})^{1-\alpha_x}}\right),
\label{eq_cum}
\end{equation}
where $\alpha_x$ is the variable parameter referring to the power-law index of the distribution, $x_0$ is a constant introduced by taking into account the threshold effect (e.g., incomplete sampling below $x_0$, background contamination), $N_{\rm env}$ is the total number of events, $x_1$ and $x_2$ are the minimum and maximum values of $x$, respectively. The uncertainty of the cumulative distribution in a given bin $i$ is approximately calculated as $\sigma_{\mathrm{cum}, i}=\sqrt{N_{cum,i}}$, where $N_{cum,i}$ is the number of events in the $i$th bin.

The standard reduced chi-square ($\chi_{\nu}^{2}$) goodness is used to confirmed a best fit. The $\chi_{\nu}$ can
be expressed as
\begin{equation}
\chi_{\nu,cum}=\sqrt{\frac{1}{(n_x-n_{par})}\sum_{i=1}^{n_x}\frac{\left[N_{cum,th}(x_i)-N_{cum,obs}(x_i)\right]^2}{\sigma_{cum,i}^2}}
\label{}
\end{equation}
for the cumulative distribution function \citep{2015ApJ...814...19A}, where $n_{\mathrm{x}}\:$ is the number of logarithmic bins, $n_{\mathrm{par}}\:$ is the number of the free parameters, $N_{\mathrm{cum,obs}}(x_{i})\:$ is the observed values, $N_{\mathrm{cum,th}}(x_{i})\:$ is the corresponding theoretical values for cumulative distribution, respectively. It should be noted that the points below the threshold $x_0$ are just noise and do not contribute to the accuracy of the best fitting of the power-law exponentials, so they are ignored when calculating the reduced chi-square.

Owing to the limited number of bursts, we just analyzes a cumulative distribution rather than a differential one.
Only two free parameters appeared in the cumulative distribution function. In general, the cumulative distributions can be generated exceed the threshold $x_0$ due to incomplete sampling of the selected samples. Taking the threshold $x_0$ as the free parameter and adding the exponent $\alpha_x$ to fit the cumulative distribution, the power-law exponent $\alpha_x$ of the samples can be well constrained. At the same time, it should be indicated that the cumulative number distribution of Eq.~(\ref{diff_cum}) is a power-law function with an exponent of $\alpha_x$, so it is an important quest of this work to deduce the power-law exponent from the data of these burst values.

In this work, the python module pymc\footnote{https://pypi.org/project/pymc/} is utilized to compile the selected data and the Monte Carlo Markov chain (MCMC) method is used to obtain the confidence intervals of the fitting parameters. Due to oversimplified sampling at a low value threshold, the distributions of the selected parameters usually show a shallow part or a gap before the threshold $x_0$ (\citealp{2015ApJ...814...19A,2015ApJS..216....8W}).

\section{Results and Discussion}
\label{sect:discuss}

\begin{table}[ht]%
\centering
\caption{The best-fitting power-law indices ($\alpha_x$) of these parameters.}
\begin{tabular}{ccccccc}
\hline
\hline
Band       & \multicolumn{1}{c}{Net Counts} & \multicolumn{1}{c}{Duration} & \multicolumn{1}{c}{Waiting Time} \\
\hline
Total(1-250 keV)       & 1.63 $\pm$ 0.08        & 1.99 $\pm$ 0.08       & 1.95 $\pm$ 0.08           \\
\hline
HE(28-250 keV)         & 1.57 $\pm$ 0.03        & 2.02 $\pm$ 0.18       & 1.95 $\pm$ 0.08           \\
ME(10-30 keV)          & 1.58 $\pm$ 0.03        & 1.99 $\pm$ 0.11       & 1.95 $\pm$ 0.07           \\
LE(1-10 keV)           & 1.58 $\pm$ 0.03        & 2.07 $\pm$ 0.15       & 1.97 $\pm$ 0.10           \\
\hline
\end{tabular}
\end{table}

The scale-free power-law distribution of various events for different parameters, such as, duration, energy or luminosity, is one of the obvious characteristics of SOC systems \citep{2014MNRAS.439.3439P,2022RAA....22f5012Z}. According to the theoretical framework proposed by \cite{2012A&A...539A...2A}, the concept of the fractal dimensions of an SOC avalanche systems can be quantitatively connected to the cumulative frequency distributions. In this framework, the power law index of SOC cumulative frequency distribution for different parameters is theoretically associated with the Euclidean space dimension S = 1, 2, 3.

\cite{2012A&A...539A...2A} reported theoretical indices such as the duration frequency distribution ($\alpha_{T}$) and the distribution of energy ($\alpha_{E}$), which aligning with $\alpha_{T}=\frac{S+1}{2}$ and $\alpha_{E}=\frac{3(S+1)}{S+5}$ , where $S=1$, 2 and 3 are the Euclidean dimensions. We can see clearly that the indices are $\alpha_{T}$=$\alpha_{E}=1$ for $S=1$, $\alpha_{E}=1.29$ and $\alpha_{T}=1.5$ for $S=2$, $\alpha_{T}=2$ and $\alpha_{E}=1.5$ for $S=3$, respectively. Here, the net counts reflect the amount of released energy of the X-ray bursts.

In order to check the possible evolution across different energy bands, the distributions of three parameters in HE, ME, LE and the total energy bands are analysed and shown in Figure 1. The detailed fitting results are listed in Table 1. As shown in Table 1 and Figure 1, the best-fitting power-law indexed of the cumulative distribution for the net photon counts in the total energy band, HE, ME and LE are 1.63$\pm$0.08, 1.57$\pm$0.03, 1.58$\pm$0.03 and 1.58$\pm$0.03, which are basically consistent with the theoretical index $\alpha_{E}=1.5$ for $S=3$. The mean value of these indexes of four individual channels is 1.59$\pm$0.02, which is larger than those of the fluence reported by \cite{2022ApJS..260...25C}. The steeper index
of the net photon count distributions may be due to the difference of the range of the X-ray burst fluence. Similarly, the results of the duration and the waiting time in the four energy bands are 1.99$\pm$0.08, 2.02$\pm$0.18, 1.99$\pm$0.11, 2.07$\pm$0.15 and 1.95$\pm$0.08, 1.95$\pm$0.08, 1.95$\pm$0.07, 1.97$\pm$0.10, and also well consistent with the theoretical index $\alpha_{T}=2$ for $S=3$.

In addition, it can be found that there is no obvious power-law index evolution across different energy bands, which gives a sufficient evidence to the SOC behavior in X-ray bursts produced by SGR J1935+2154. Moreover, the Euclidean space dimension S=3 can be obtained according to the prediction of the FD-SOC model. It is the key point that the relationship among the indices of power-law distributions of different parameters of X-ray bursts is mainly dependent on the nonlinear scaling law among the self-organized critical parameters.

Much previous studies have been done on the statistical properties of SGRs, most of which focus on the distribution of burst energy\citep[e.g.][]{1996Natur.382..518C,1999ApJ...526L..93G,2000ApJ...532L.121G,2020MNRAS.491.1498C}. For example,
\cite{1996Natur.382..518C} found that SGR events and earthquakes share four unique statistical properties, including but not limited to power-law energy distribution and log-symmetric waiting time distribution. These statistical similarities suggest that SGRs should be powered by star-quakes like earthquakes.

\cite{1999ApJ...526L..93G} conducted a similar study of the bursts from SGR 1900+14 during its 1998-1999 active phase. They found that the distribution of fluence or energy of SGR 1900+14 is aligning with a power law index of 1.66 and firstly reported the SOC behaviors in SGR bursts. Then, they further
presented the statistics of bursts from SGR 1806-20 detected by the Rossi X-Ray Timing Explorer/Proportional Counter Array, the Burst and Transient Source Experiment, and the International Cometary Explorer \citep{2000ApJ...532L.121G}. They found that the distribution of bursts' fluence observed with each instrument are well characterized by power laws with the indices 1.43, 1.76, and 1.67, respectively.
They proposed the hypothesis that the source of energy for the SGR bursts may not any accretion or nuclear power, but rather crustquakes caused by the neutron star's evolving strong magnetic field.

Similarly, the total counts or energy of SGR bursts are verified to have power-law-like size distributions. \cite{2020MNRAS.491.1498C} found that the energy distribution of magnetar bursts can be well described by power-law functions with exponents of 1.84, 1.68, and 1.65 for the three events of SGR J1550-5418, SGR 1806-20, and SGR 1900+14, and the duration distributions of them are also show power-law forms with exponents of 1.69, 1.72 and 1.82, respectively. Meanwhile, the distribution of waiting time can be described by a non-stationary Poisson process in which the occurrence rate increases exponentially. In Tables 2 and 3, we present a comparison of present relevant studies about SGRs, including the results of SGR J1935+2154 and the other SGRs detected by different instruments.

\cite{2023RAA....23c5007L} studied the energy spectra of these 75 bursts using a variety of models and found that the spectra of magnetar bursts are complex and diverse, and different types of bursts exhibit unique characteristics on the phase distributions due to different generation mechanisms. However, they found that most of the optimal energy spectrum can be fitted by the cutoff power-law and power-law models, which coincides with the power-law distribution on the statistical characteristics of the parameters.

The SOC features have been found in FRB 121102 \citep[e.g.][]{2020MNRAS.491.1498C,2021ApJ...920L..23Z}. Recently, \cite{2023arXiv230206802W} studied the frequency distributions of burst energy and waiting time of the two repeating FRBs 20121102A and 20201124. According to the bimodal distributions of the waiting time, the bursts are divided into long and short parts. It is found that the two characteristics of both long and short bursts can be fitted by a power-law function and well understood within the physical framework of an SOC system driven in a correlated way.
They propose a possible trigger mechanism that favors the emission of the star's magnetosphere, the crustal failure of a neutron star. The detection of FRB 200428 confirmed that at least some of the FRBs originate from magnetars \citep{2020Natur.587...54C,2020Natur.587...59B}. \cite{2021ApJ...920..153W} studied the properties of X-ray bursts from SGR J1935+2154 observed by the Gamma-ray Burst Monitor (GBM) on board and FRB 121102 detected by Five-hundred-meter Aperture Spherical radio Telescope.
They found that the high-energy components of FRB 121102 and SGRs have similar scale-invariant behavior, which can be well interpreted by the same SOC framework with spatial dimension $S = 3$.

Regarding the generation mechanism of magnetar X-ray bursts, \cite{2022ApJ...933..174Y} simulated the three-dimensional force-free electrodynamics of local Alfv$\acute{e}$n wave packets emitted by magnetar vibrations into the magnetosphere. They found that if the Alfv$\acute{e}$n wave packet propagates to a radius R and the total energy is greater than a certain value of magnetosphere energy, the wave would become very nonlinear and be ejected from the magnetosphere. The ejecta can carry a large energy of the initial Alfv$\acute{e}$n wave. Then it opens up the magnetospheric magnetic field lines, forming a current sheet behind them that connects back to the enclosed area. Magnetic reconnection will occur on these current sheets, which will cause plasma excitation and X-ray emission. We believe that the SOC process may occur during this period.

In this work, we further studied the properties of X-ray bursts from SGR J1935+2154 observed by HXMT and found the similar evidences of SOC power-law behaviors in all different energy channels. Therefore, we suggest that X-ray bursts from SGR J1935+2154 can be possibly considered as ``avalanches" in SOC systems, in which the magnetic reconnection occurs during these X-ray bursts.

\section{Conclusions}

In this work, we have statistically analyzed three parameters of 75 X-ray bursts produced by SGR J1935+2154 during the active period beginning on 27 April 2020, including durations, waiting times, and net photon counts. We checked the cumulative distribution of these parameters on different energy bands. We found that all the three parameters of X-ray bursts have similar power-law distributions, thus all can be explained by an SOC behavior. Moreover, it is found that there is no obvious power-law index evolution among different energy channels. In addition, the Euclidean space dimension $S=3$ has been obtained. Above all, we have obtained the sufficient evidence that the X-ray bursts arise from a mechanism dominated by self-organizing critical systems, which will help us to further study the radiation mechanism of magnetars.

\section*{Acknowledgements}
\addcontentsline{toc}{section}{Acknowledgements}

This work is supported by National Key R\&D Program of China (2021YFA0718500), the National Natural Science Foundation
of China under grants U2038106 and 12065017, and partially by the Jiangxi Provincial Natural Science Foundation under grant 20224ACB211001.

\begin{figure*}[ht!]
\includegraphics[width=0.33\textwidth, angle=0]{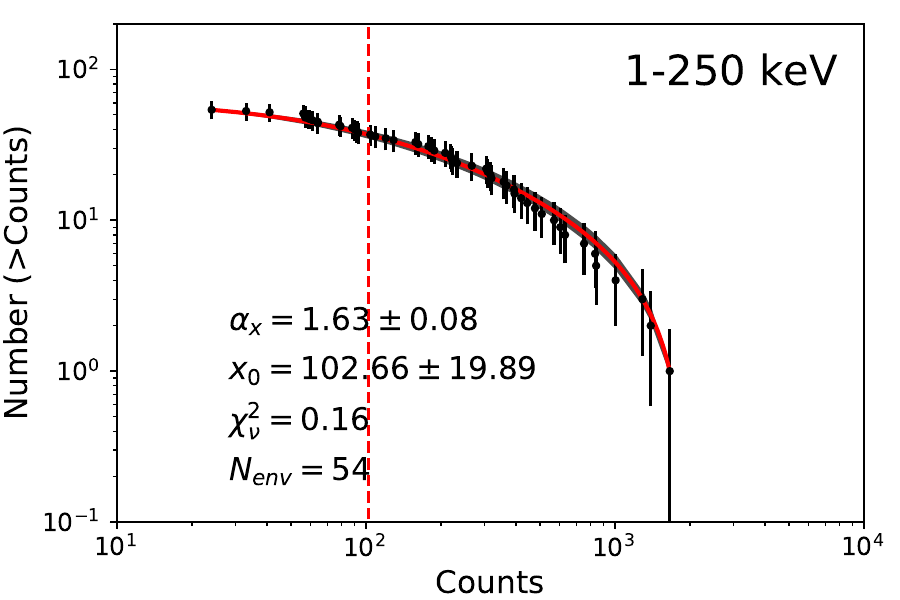}\includegraphics[width=0.33\textwidth, angle=0]{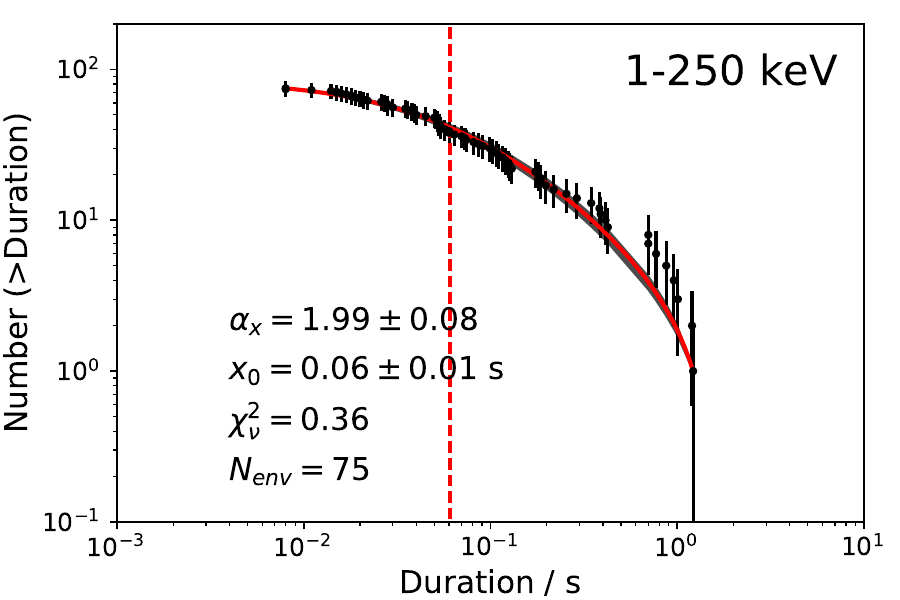}\includegraphics[width=0.33\textwidth, angle=0]{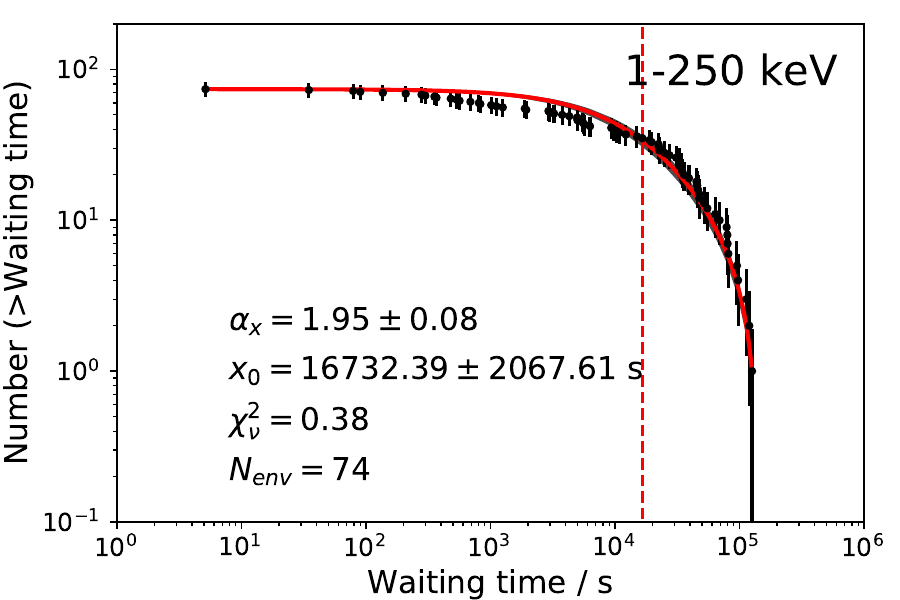}\\
\includegraphics[width=0.33\textwidth, angle=0]{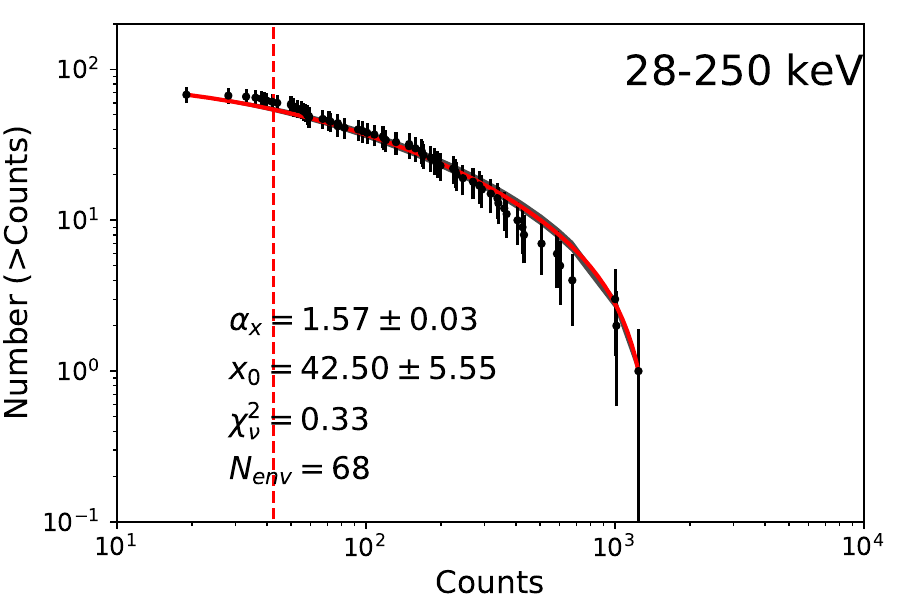}\includegraphics[width=0.33\textwidth, angle=0]{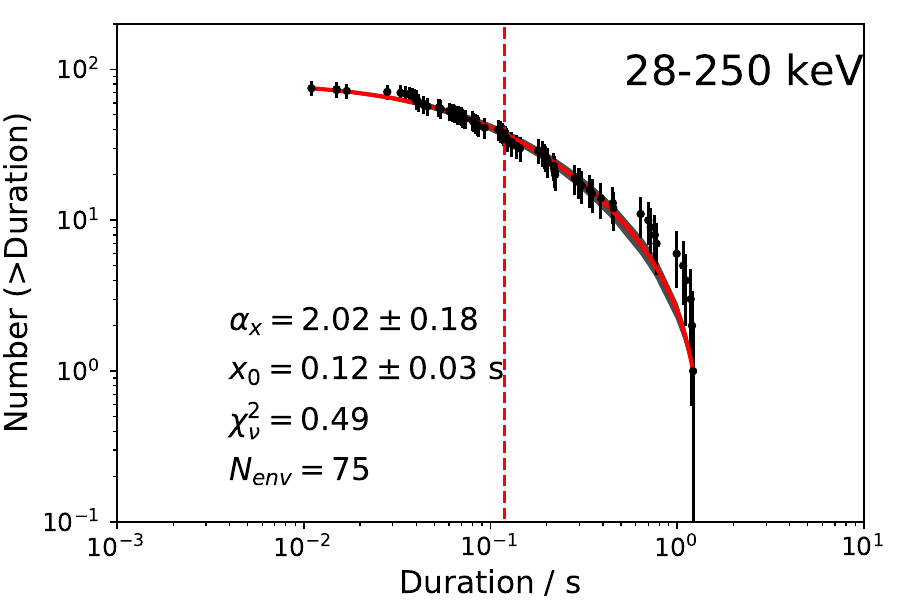}\includegraphics[width=0.33\textwidth, angle=0]{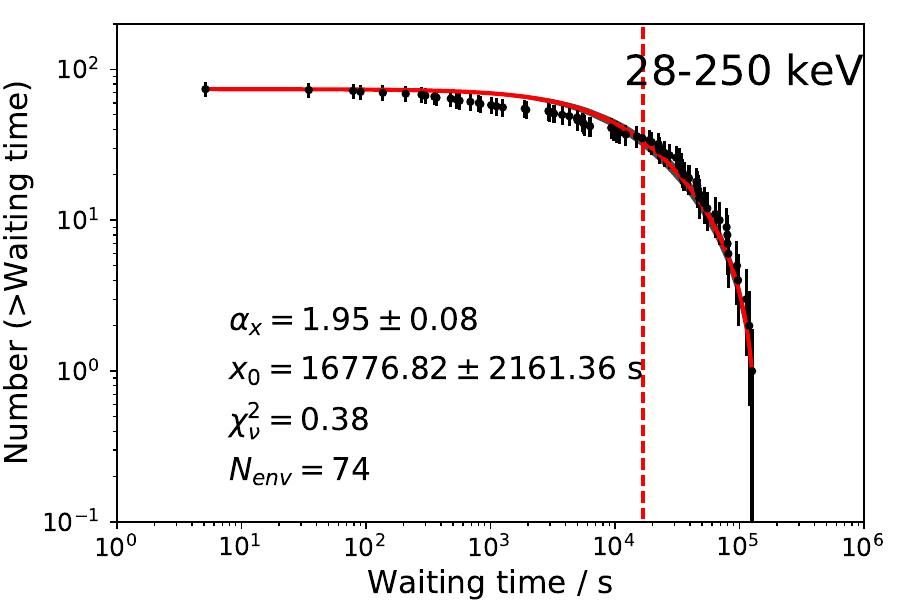}\\
\includegraphics[width=0.33\textwidth, angle=0]{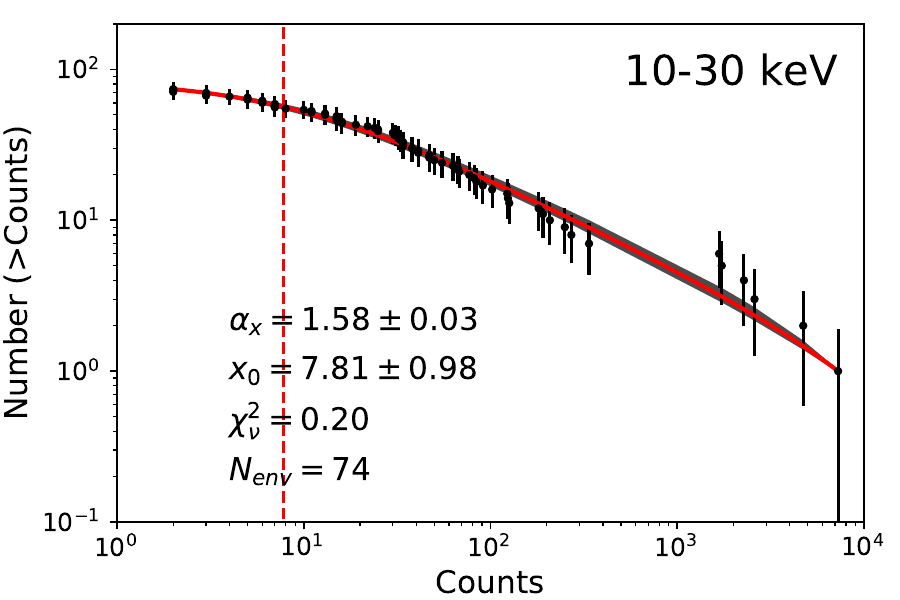}\includegraphics[width=0.33\textwidth, angle=0]{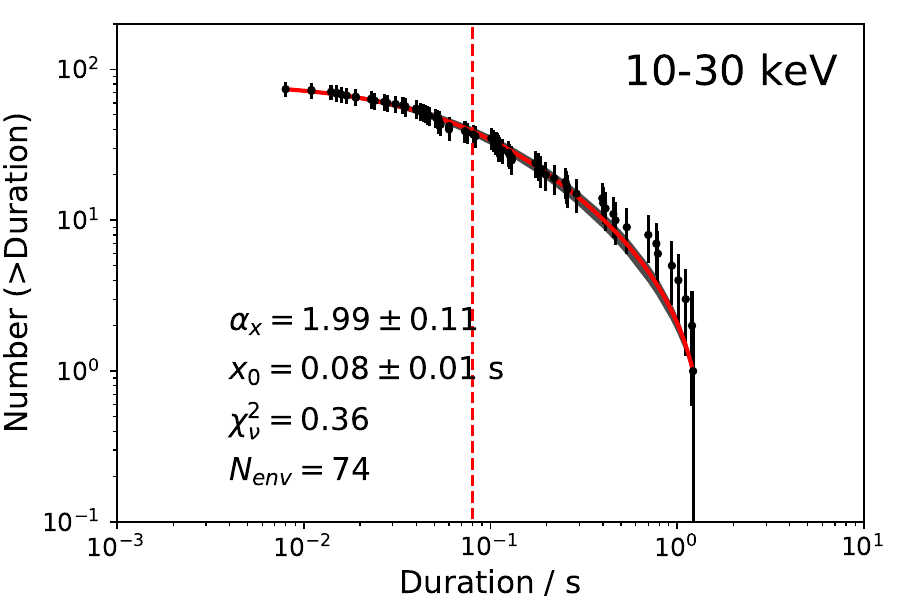}\includegraphics[width=0.33\textwidth, angle=0]{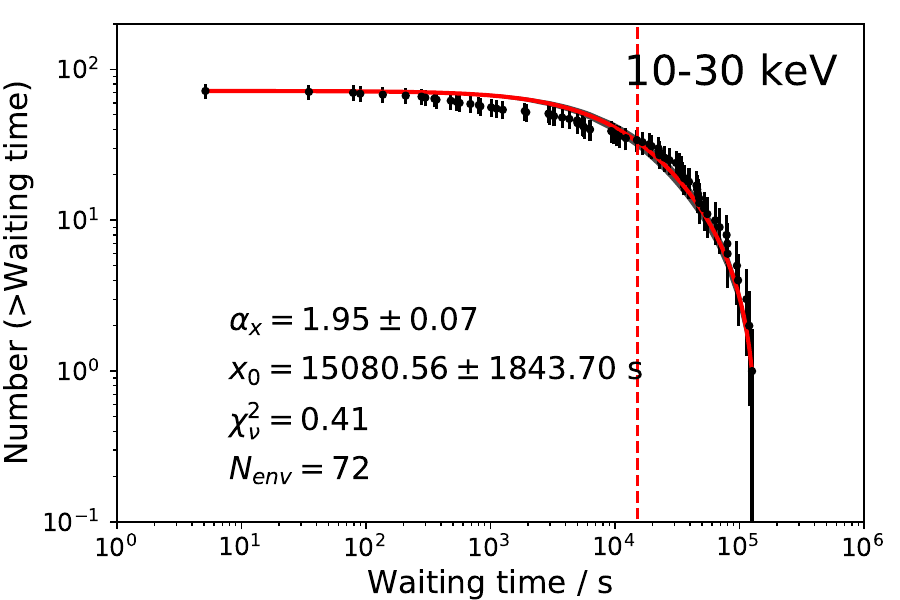}\\
\includegraphics[width=0.33\textwidth, angle=0]{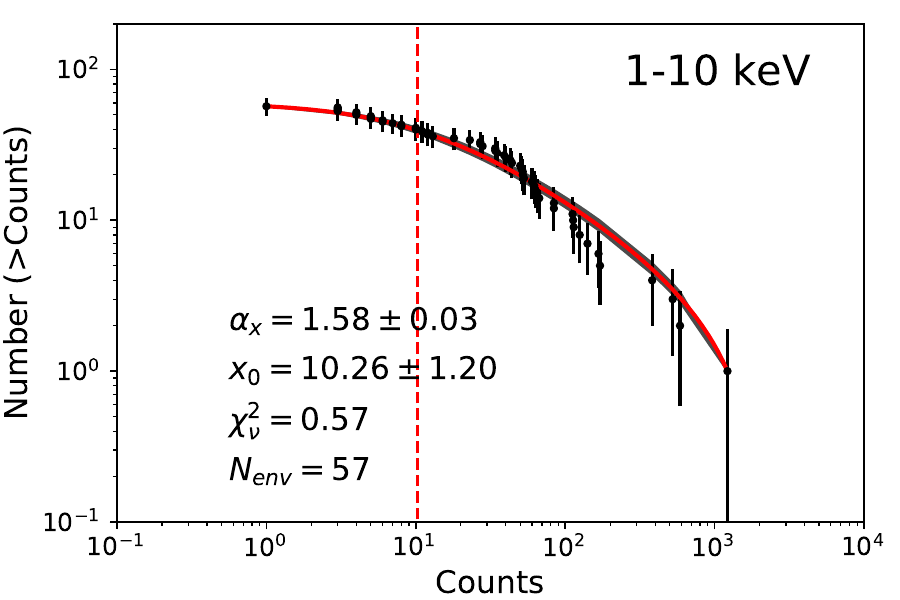}\includegraphics[width=0.33\textwidth, angle=0]{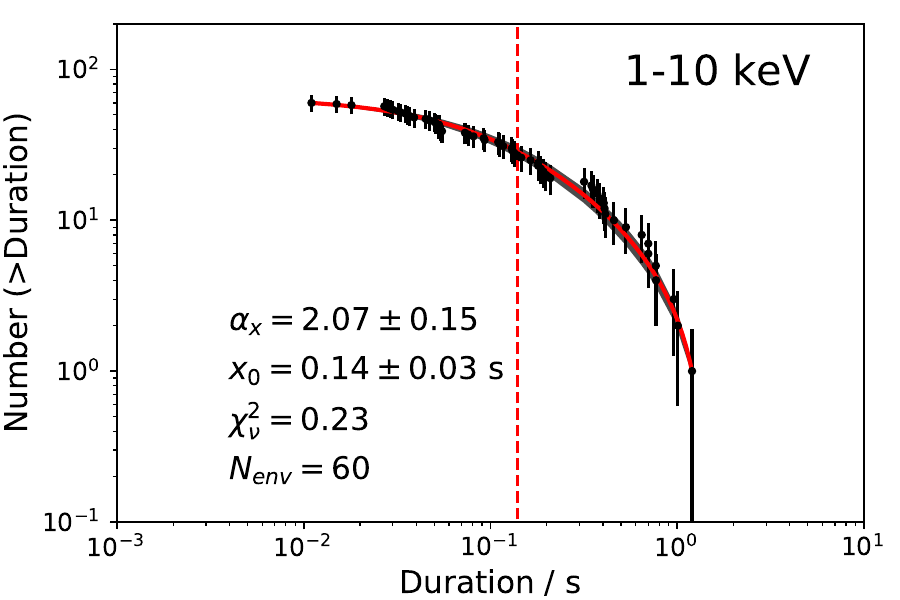}\includegraphics[width=0.33\textwidth, angle=0]{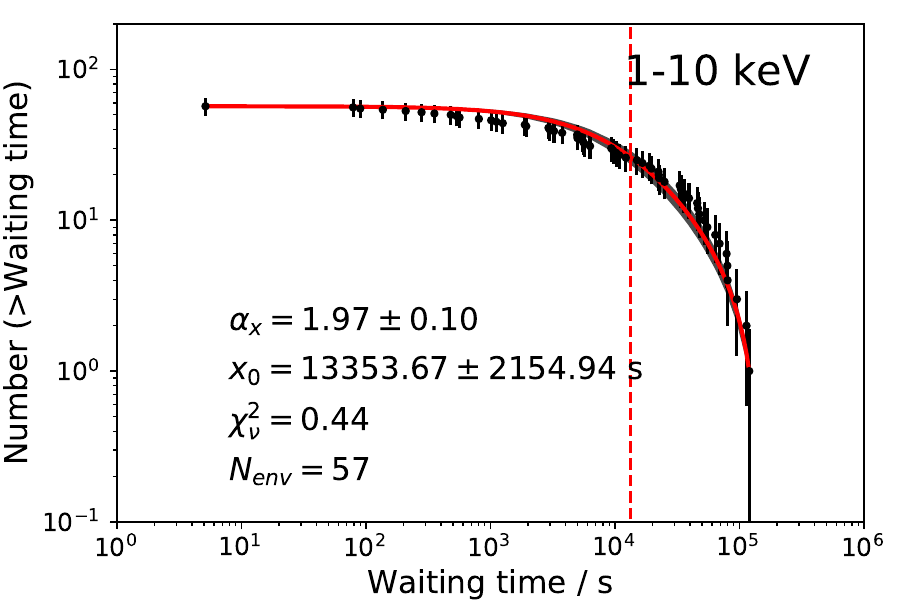}\\
\caption{The cumulative distributions of the net count, duration and waiting time of the X-ray bursts from SGR J1935+2154. The gray region represents the 95$\%$ confidence level, the red line is the best fitting result, and the dashed line is marked as the threshold $x_0$.
Note that the numbers of data of each parameter adopted have been marked in each panel. }
\label{fig:flare}
\end{figure*}

\begin{table}[ht!] 
\centering
\caption{Results of some previous studies of SGRs.}
\begin{threeparttable}
\begin{tabular}{ccccccc}
\hline
\hline

\multicolumn{1}{c|}{SGR}                             & Distribution\tnote{a} & Function\tnote{b} & $\alpha_T$\tnote{c} & $\alpha_E$\tnote{d} & Reference        \\ \hline
\multicolumn{1}{c|}{SGR 1627-41}                     & cumulative   &  $PL$   & ... &0.62$\pm$0.08&\cite{1999ApJ...519L.139W}\\ \hline
\multicolumn{1}{c|}{SGR J1550-5418}                  & cumulative   &  $PL$   & 1.69$\pm$0.03 &1.84$\pm$0.03&\cite{2020MNRAS.491.1498C}\\ \hline
\multicolumn{1}{c|}{\multirow{2}{*}{SGR 1806-20}}    & differential &  $PL$   & ... &1.6&\cite{1996Natur.382..518C}  \\
\multicolumn{1}{c|}{}                                & cumulative   &  $Threshold-PL$   &1.72$\pm$0.01&1.68$\pm$0.01&\cite{2020MNRAS.491.1498C}\\ \hline
\multicolumn{1}{c|}{\multirow{2}{*}{SGR 1900+14}}    & differential &  $PL$   & ...&1.66$\pm$0.05&\cite{1999ApJ...526L..93G}\\
\multicolumn{1}{c|}{}                                & cumulative   &  $Threshold-PL$    &1.82$\pm$0.02&1.65$\pm$0.01&\cite{2020MNRAS.491.1498C}\\ \hline
\end{tabular}

 \begin{tablenotes}
        \footnotesize
        \item[a] The purpose of distinguishing the forms of distributions is that because they are integral and differential to each other, the difference between the results of the same function fits is 1.  
        \item[b] $PL$ is a simple power law function($F\propto x^{-\alpha_x}$), and $Threshold-PL$ is a threshold power law function of the two distribution forms used in this work. 
        \item[c] The time-dependent parameters of a burst, including duration, waiting time, etc.
        \item[d] The energy-related parameters of a burst, including energy, fluence, flux, photon-counts, etc.
      \end{tablenotes}
    \end{threeparttable}

\end{table}

\begin{table}[ht!]
\centering
\caption{Comparison of observation results of the SGRs by different instruments.}
\begin{tabular}{ccccccc}
\hline
\hline
\multicolumn{1}{c|}{SGR}                             & Distribution & Function & $\alpha_E$  & \multicolumn{2}{c}{Instrument}   & Reference \\ \hline
\multicolumn{1}{c|}{\multirow{3}{*}{SGR J1935+2154}} & differential &    $PL$    & 1.55$\pm$0.01 & \multicolumn{2}{c}{NICER}        & \cite{2020ApJ...904L..21Y} \\
\multicolumn{1}{c|}{}                                & cumulative   &    $PL$    & 0.93$\pm$0.15& \multicolumn{2}{c}{Fermi-GBM}    & \cite{2023ApJ...950..121R}\\
\multicolumn{1}{c|}{}                                & cumulative   & $Threshold-PL$  & 1.63 $\pm$ 0.08 & \multicolumn{2}{c}{Insight-HXMT} & this work       \\ \hline
\multicolumn{1}{c|}{\multirow{3}{*}{SGR 1806-20}} & differential & $PL$ & 1.43$\pm$0.06& \multicolumn{2}{c}{PCA}        & \multirow{3}{*}{\cite{2000ApJ...532L.121G}} \\
\multicolumn{1}{c|}{}                                & differential   &$PL$ &1.76$\pm$0.17& \multicolumn{2}{c}{BATSE}    &                   \\
\multicolumn{1}{c|}{}                                & differential   &$PL$&1.67$\pm$0.15& \multicolumn{2}{c}{ICE} &                   \\ \hline
\end{tabular}
\end{table}

\end{document}